# Machine-learning-assisted thin-film growth: Bayesian optimization in molecular beam epitaxy of SrRuO$_3$ thin films

Yuki K. Wakabayashi,[1,a] Takuma Otsuka,[2] Yoshiharu Krockenberger,[1] Hiroshi Sawada,[2] Yoshitaka Taniyasu,[1] and Hideki Yamamoto[1]

[1]*NTT Basic Research Laboratories, NTT Corporation, Atsugi, Kanagawa 243-0198, Japan*
[2]*NTT Communication Science Laboratories, NTT Corporation, Soraku-gun, Kyoto 619-0237, Japan*
a)Author to whom correspondence should be addressed: yuuki.wakabayashi.we@hco.ntt.co.jp

Abstract
Materials informatics exploiting machine learning techniques, *e.g.*, Bayesian optimization (BO), has the potential to offer high-throughput optimization of thin-film growth conditions through incremental updates of machine learning models in accordance with newly measured data. Here, we demonstrated BO-based molecular beam epitaxy (MBE) of SrRuO$_3$, one of the most-intensively studied materials in the research field of oxide electronics, mainly owing to its unique nature as a ferromagnetic metal. To simplify the intricate search space of entangled growth conditions, we ran the BO for a single condition while keeping the other conditions fixed. As a result, high-crystalline-quality SrRuO$_3$ film exhibiting a high residual resistivity ratio (RRR) of over 50 as well as strong perpendicular magnetic anisotropy was developed in only 24 MBE growth runs in which the Ru flux rate, growth temperature, and O$_3$-nozzle-to-substrate distance were optimized. Our BO-based search method provides an efficient experimental design that is not as dependent on the experience and skills of individual researchers, and it reduces experimental time and cost, which will accelerate materials research.

# I. INTRODUCTION

The itinerant ferromagnetic perovskite SrRuO3 is one of the most promising materials for oxide electronics.[1-10] Owing to its compatibility with other perovskite-structured oxides, as well as its high conductivity and chemical stability,[3] SrRuO3 is widely used as an epitaxial conducting layer in oxide heterostructures. However, a thorough understanding of its transport properties, electronic structure, and origin of its ferromagnetism remains elusive despite tremendous efforts for over five decades. While high-quality specimens are indispensable for exploring electronic states, it is difficult to make high-quality bulk single crystals of SrRuO3, and hence, thin film specimens have been making a significant contribution to such research. The residual resistivity ratio (RRR), which is defined as the ratio of resistivity at 300 K [$\rho(300\ K)$] to that at 4 K [$\rho(4\ K)$], is a good measure of the purity of a metallic system, and accordingly, the quality of single-crystalline SrRuO3 thin films: RRR is very sensitive to defects and off-stoichiometry.[8,11,12] More specifically, only SrRuO3 films with high RRR values above 40 and 60 have enabled observation of sharp dispersive quasiparticle peaks near the Fermi level by angle-resolved photoemission spectroscopy (ARPES) and of quantum oscillations in the electrical resistivity, respectively.[13,14] Such high-quality films maintain their metallic and ferromagnetic character even when the thickness is reduced to a monolayer,[15] providing a rare example of two-dimensional ferromagnetism. This means that extremely thin SrRuO3 films can serve as a two-dimensional spin-polarized electron system, as the existence of a spin-polarized electron current has been established in thicker SrRuO3-based magnetic tunnel junctions.[16,17] Accordingly, high-quality SrRuO3 thin films are also promising for future spintronics applications. However, fine-tuning of multiple growth conditions, including the flux ratio of Ru/Sr, the growth temperature, and the oxygen pressure, is required for high-RRR SrRuO3 growth, and only a few papers have reported SrRuO3 films with RRRs of over 50.[7,10,14]

To optimize the growth conditions, one may take a conventional trial-and-error approach, which is time-consuming and costly, and whose optimization efficiency largely depends on the skills and experience of individual researchers. In contrast, recent advances in materials informatics exploiting machine learning techniques, such as Bayesian optimization (BO) and artificial neural networks, offer an alternative approach of high-throughput experiments.[18-22] So far, materials informatics has mainly focused on high-throughput predictions of brand-new materials having designated functions by utilizing materials databases and/or theoretical calculations.[22-30] Nevertheless, there have been a few reports on high-throughput experiments achieved by incrementally updating the machine learning models in accordance with newly measured data.[31-35] Since BO is a sample-efficient approach for global optimization,[36] adaptive sampling of the growth conditions with BO will streamline the optimization of the thin-film growth conditions.

In this article, we describe machine-learning-assisted molecular beam epitaxy (MBE) of the SrRuO3 thin films. We developed a high RRR (> 50) SrRuO3 film in only 24 MBE growth runs. Thorough crystallographic analyses confirmed the high crystallinity of the films with the large RRR values. In addition, these films showed perpendicular magnetization with rectangular hysteresis and a small coercive field of 0.1 T, which is advantageous for spintronics applications. The algorithm presented here is an efficient means for experimental designs aimed at optimizing the growth conditions, with which advancement will be less empirical as compared with the conventional process and at significant reductions in experimental time and cost.

## II. EXPERIMENT

Single-crystalline SrRuO$_3$ thin films (65-nm thick) were epitaxially grown on (001) SrTiO$_3$ substrates in a custom-designed MBE setup equipped with multiple e-beam evaporators for Sr and Ru.[37-39] We precisely controlled the elemental fluxes even for elements with high melting points, *e.g.*, Ru (2250°C), by monitoring the flux rates with an electron-impact-emission-spectroscopy sensor and feeding the results back to the power supplies for the e-beam evaporators. The Sr flux rate was fixed to 0.98 Å/s. The oxidation during growth was carried out with ozone (O$_3$) gas. The O$_3$ gas (~15% O$_3$ + 85% O$_2$) was introduced through an alumina nozzle pointed at the substrate at a flow rate of ~2 sccm. The nozzle-to-substrate distance is an important growth parameter that, along with the growth temperature, determines the local oxidation strength at the growth surface. Further information about the MBE setup and preparation of the substrates are described elsewhere.[39] In the conventional synthesis procedure, three important growth parameters; *i.e.*, the Ru flux rate, growth temperature, and O$_3$-nozzle-to-substrate distance, are systematically optimized, but in an empirical manner whose success depends on the experience of the researchers. In contrast, we optimized the three parameters by using the BO algorithm. The Ru flux rate, growth temperature, and nozzle-to-substrate distance were varied in ranges between 0.18 and 0.61 Å/s, 565 and 815°C, and 1 and 31 mm, in correspondence with the search ranges in BO. Here, the search range of the growth temperature is within the typical thermodynamic growth window for growth of SrRuO$_3$ thin films.[10] We searched equally spaced grid points for each parameter. The number and corresponding intervals of the respective quantities were 44 (0.01 Å/s interval), 41 (6.25°C interval), and 31 (1 mm interval). Since the three-dimensional parameter space consisted of 55924 (44×41×31) points, full surveillance of the entire space in a point-by-point manner was unrealistic, as only a limited number runs can be carried out per day with a typical MBE system. The crystal quality of the films was monitored by *in-situ* reflection high-energy electron diffraction (RHEED) during and after the growth. When diffractions from the SrRuO$_3$ phase were indiscernible and/or diffractions from SrO or RuO$_2$ precipitates appeared, we defined the RRR value of those samples to be 0.[10]

Figure 1 shows the procedure of machine-learning-assisted MBE growth of the SrRuO$_3$ thin films based on the BO algorithm, which is a sequential design to optimize a black-box function S($x$).[40] In this study, we optimized the growth parameters one dimension at a time. In concrete terms, we first optimized the Ru flux rate while keeping the other parameters unaltered. Subsequently, we tuned the growth temperature and the nozzle-to-substrate distance. Here, RRR = S($x$) is the target function specific to our SrRuO$_3$ films, and $x$ is the growth parameter (Ru flux rate, growth temperature, or nozzle-to-substrate distance). Given a data set $\{x_m, \text{RRR}_m\}_{m=1}^{M}$ obtained from the past $M$ MBE growths and RRR measurements of SrRuO$_3$ films, we use it to construct a model to predict the value of S($x$) at an unseen $x$. To this end, we use Gaussian process regression (GPR) to estimate the mean ($\mu$) and variance ($\sigma$) at an arbitrary parameter value $x$. Specifically, GPR predicts S($x$) that follows a Gaussian distribution N($\mu$, $\sigma^2$), where $\mu$ and $\sigma$ depend on $x$ and the $M$ data samples. In short, $\mu(x)$ represents the expected value of RRR and $\sigma(x)$ represents the uncertainty of RRR at $x$. To take into account the inherent noise in the RRR of SrRuO$_3$ films grown under nominally the same conditions, the variance of the observation noise $\beta$ of the GPR model was set to 0.02 or 0.002. Further information about the GPR is described elsewhere.[33] We iterate the routine after the initial

MBE growth and RRR measurements for each growth parameter optimization. First, GPR is updated using the data set at the time of optimization for the specific parameter currently being surveyed. Subsequently, to assign the value of the growth parameter in the next run, we calculate the expected improvement (EI);[41] EI balances exploitation and exploration by using the predicted $\mu(x)$, $\sigma(x)$, and the best experimental RRR value at that time. More specifically, an unmeasured value of $x$ at which the EI($x$) takes a maximum is sampled in the next sequence. Note that we excluded already measured $x$ values from being the next sampling point to avoid duplicate trials. This routine is iterated to grow and measure more samples until an arbitrary criterion is satisfied. Examples of such criteria include the degree of convergence of EI values and setting a predetermined number of sampling points. Here, we stopped the routine at 11 samples per parameter. After completing 11 samplings for a certain parameter, we chose the value that gave the highest RRR and started the optimization of another parameter.

## III. RESULTS AND DISCUSSION
### A. Bayesian optimization of MBE growth of SrRuO$_3$

First, we optimized the Ru flux rate while keeping the growth temperature and the nozzle-to-substrate distance fixed (615°C and 16 mm, respectively). Here, we could have started with any initial growth temperature and nozzle-to-substrate distance, although the selection may affect the efficiency of the optimization. Figure 2 shows how the BO algorithm predicts RRR values with unseen parameter configurations and acquires new data points: the process starts with five random initial Ru flux rates [Fig. 2(a)] and gains experimental RRR values for the updated GPR model with seven [Fig. 2(b)], nine [Fig. 2(c)], and eleven [Fig. 2(d)] Ru flux rates. The EI values at the respective stages are shown in the lower panels. In Figs. 2(b)-2(d), the search range was reduced to the Ru flux rate range within which the SrRuO$_3$ phase had formed. For the RRR predicted with five initial data points [Fig. 2(a)], the $\sigma$ value above 0.4 Å/s increased with increasing Ru flux rate. This indicates that the RRR prediction in that region is uncertain because of the absence of data points above 0.4 Å/s. The results in Fig. 2(a) indicate that the highest EI was at 0.61 Å/s. The overall variance of the predicted RRR became smaller as the data points increased from five to 11, leading to lower EI values. This suggests that we have only a limited chance to improve the RRR value by further modification of the Ru flux rate. The highest experimental RRR (29.33) was obtained at 0.42 Å/s, which was in good agreement with the prediction, and hence, we set the Ru flux rate to 0.42 Å/s in the next optimization; namely, the growth temperature optimization.

The growth temperature optimization was carried out at the optimized Ru flux rate (0.42 Å/s) and at a fixed initial (not yet optimized) nozzle-to-substrate distance (16 mm). Four initial growth temperatures were chosen at equal intervals [Fig. 3(a)]. As in the case of the Ru flux rate optimization, the variance of the predicted RRR became smaller as more data points were collected, resulting in a reduction in EI. For 721°C, the maximum experimental RRR (49.23) coincided with the prediction [Fig. 3(d)]. Accordingly, we fixed the growth temperature for the nozzle-to-substrate distance optimization at 721°C.

We optimized the O$_3$-nozzle-to-substrate distance while keeping the Ru flux rate and growth temperature at their optimized values. Three initial O$_3$-nozzle-to-substrate distances were chosen at equal intervals [Fig. 4(a)]. The search range was reduced to the nozzle-to-substrate distance range within which the SrRuO$_3$ phase had formed. Again, the variance of the predicted RRR became smaller, with a concomitant reduction in EI, as the number of data points increased. The highest experimental RRR, 51.79, was achieved at a nozzle-to-substrate distance of 15 mm. Through these optimizations, SrRuO$_3$ films with RRRs over 10, 20, 40, and 50 were obtained in four, five, 12, and 24 MBE growth runs (Fig. 5).

The quality of the SrRuO$_3$ films exhibiting RRRs over 50 was high enough to investigate the intrinsic electronic structure by, *e.g.*, ARPES, and to make an atomically thin ferromagnetic and conducting layer.[13,15] However, the highest RRR (51.79) was still lower than the highest value ever reported (~80).[7] To increase the RRR of our SrRuO$_3$ films, it is necessary to find the global-best point in the three-dimensional parameter space. In the following, we detail two methods to enhance the RRR value. One is to iterate the parameter-wise optimization while setting the other parameters to their best values; *i.e.*, to restart with the optimization of the Ru flux rate with the growth temperature and nozzle-to-substrate distance set at 721°C and 15 mm. The other is to run the BO algorithm in the three-dimensional space directly. The latter idea, however, may lead to an inefficient search because a GPR prediction in three-dimensional space requires more data points for accuracy than a GPR prediction in one-dimensional space.

**B. Crystallographic, electrical, and magnetic properties of high-quality SrRuO$_3$ films (RRR > 50)**

We experimentally characterized the high-quality SrRuO$_3$ thin films with RRRs larger than 50. First, the crystallinity of the SrRuO$_3$ films was examined by RHEED, X-ray diffraction (XRD), atomic force microscopy (AFM), and high-angle annular dark-field scanning transmission electron microscopy (HAADF-STEM). Figure 6(a) shows the RHEED pattern of a SrRuO$_3$ thin film surface, where the sharp streaky patterns with clear Kikuchi lines indicate growth of single-crystalline SrRuO$_3$ film with an atomically flat surface. The AFM observations show a surface morphology composed of flat terraces and molecular steps with a height of ~0.4 nm, corresponding to a single unit cell (u.c.) thickness for the pseudocubic lattice of SrRuO$_3$ [Fig. 6(b)]. Laue fringes in the $\theta$-$2\theta$-scanned XRD pattern [Figure 6(c)] also indicate high crystalline quality and a large coherent volume of the film. The out-of-plane lattice constant, estimated from the Nelson-Riley extrapolation method,[42] was 3.949 Å, ~0.5% larger than the pseudocubic lattice constant of the bulk specimens (3.93 Å).[8] This implies that the SrRuO$_3$ film was compressively strained due to the lattice constant of the SrTiO$_3$ substrate (3.905 Å) being ~0.6% smaller than the pseudocubic lattice constant of SrRuO$_3$. Figures 6(d)-6(f) show cross-sectional HAADF-STEM images of SrRuO$_3$ thin film taken along the [110] direction of the SrTiO$_3$ substrate. At a glance, one can recognize that a single-crystalline SrRuO$_3$ film with an abrupt substrate/film interface had grown epitaxially on a (001) SrTiO$_3$ substrate. As the intensity in the HAADF-STEM image is proportional to ~$Z^n$ (n ~ 1.7-2.0, and $Z$ is the atomic number), the clear and periodical contrast between the cations [Sr ($Z$ = 38) and Ru ($Z$ = 44)] shown in Fig. 6(f) means the atomic arrangement in the film was uniform.

Next, the electrical and magnetic properties were investigated. The resistivity versus temperature curve shows a clear kink at ~152 K [Fig. 7(a)], at which the ferromagnetic transition occurs and spin-dependent scattering is suppressed;[7] the Curie temperature ($T_C$) of the film was ~ 152 K. A clear ferromagnetic transition at ~152 K was also observed in the magnetization versus temperature curves when a magnetic field of 100 Oe was applied to the out-of-plane [001] or in-plane [001] direction of the SrTiO$_3$ substrate [Fig.

7(a)]. $T_C$ was lower by ~ 8 K than the value reported for bulk specimens.[8] This discrepancy most likely stems from compressive strain, as the highest $T_C$ ever reported for thin film specimens epitaxially grown on SrTiO$_3$ substrates is ~152 K,[7] identical to ours. Since Ru vacancies are known to reduce $T_C$,[43] the high $T_C$ value confirms that our films were free from Ru deficiencies. Moreover, the low residual resistivity of 2.70 $\mu\Omega\cdot$cm at 4 K means that there were few defects and little off-stoichiometry, which contributed to the high RRR.[8,11,12]

Figure 7(b) plots the magnetization versus magnetic field curves at 10 K with a magnetic field applied to the out-of-plane [001] or in-plane [100] direction of the SrTiO$_3$ substrate. The saturation magnetization along the out-of-plane [001] direction was 1.25 $\mu_B$/Ru, which is a typical value for bulk and thin-film specimens.[8] In contrast, the magnetization along the in-plane [100] direction did not saturate even at 4 T. This indicates that the easy direction of magnetization was perpendicular to the film surface, as is the case with compressively strained SrRuO$_3$ films on SrTiO$_3$ substrates.[7,44,45] Perpendicular magnetic anisotropy is more advantageous than in-plane magnetic anisotropy for spintronics applications; the magnetic configuration is more thermally stable and the spin-transfer switching current is lower.[46,47] Notably, the ratio of the magnetization along the out-of-plane [001] direction at 2 T to that along the in-plane [100] direction is 2.9, which is about twice as large as those reported for SrRuO$_3$ films (1.4-1.7) on SrTiO$_3$.[7,45] Since magnetic anisotropy in magnetic oxides is often reduced by grain boundaries and other defects,[48-50] the strong perpendicular magnetic anisotropy is a hallmark of the excellent crystallinity of our SrRuO$_3$ films. In addition, as shown in Fig. 7(b), the coercive field $H_c$ was ~0.1 T along both the easy and hard axes. This value is smaller than those previously reported for SrRuO$_3$ films ($H_c > 0.2$ T).[7,8,45] Since the magnetic domains tends to be pinned by grain boundaries and other defects, the small $H_c$ would also have stemmed from the higher crystallinity.[7]

Altogether, the above results indicate that these SrRuO$_3$ thin films prepared by machine-learning-assisted MBE had high crystalline quality and strong perpendicular magnetic anisotropy.

**IV. SUMMARY**

We demonstrated machine-learning-assisted MBE growth utilizing the BO algorithm for optimization of the growth parameters and grew SrRuO$_3$ films whose RRR values were over 50. SrRuO$_3$ films with RRRs over 10, 20, 40, and 50 were developed in four, five, 12, and 24 MBE growth runs. The SrRuO$_3$ films grown under the optimized conditions showed high crystalline quality and strong perpendicular magnetic anisotropy. The algorithm presented here will play an important role in thin-film growth of various materials, as it offers an efficient experimental design platform that is not as dependent on the experience and skills of individual researchers. Since the algorithm provides the growth condition that should be examined in the next run, we foresee that the optimization process can be automated to work in combination with automatic thin-film growth[51-53] and characterization systems.

**ACKNOWLEDGEMENT**
We thank Ai Ikeda for her help with the maintenance of the MBE system.

**Figures and figure captions**

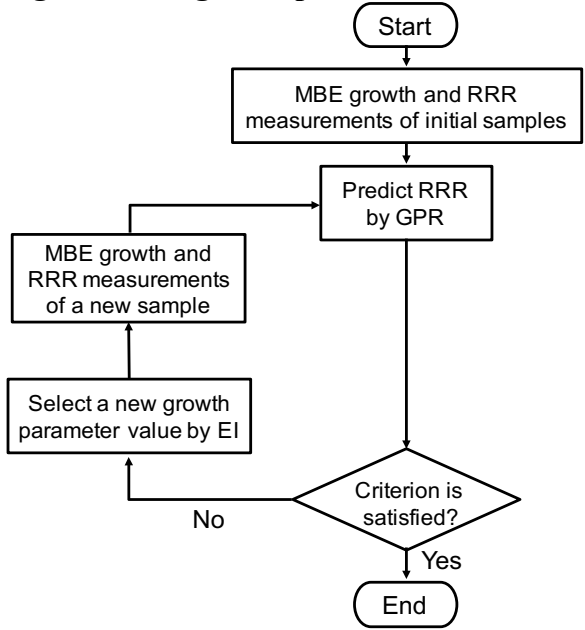

**FIG. 1.** Flowchart of machine-learning-assisted MBE growth of SrRuO$_3$ thin films based on the BO algorithm.

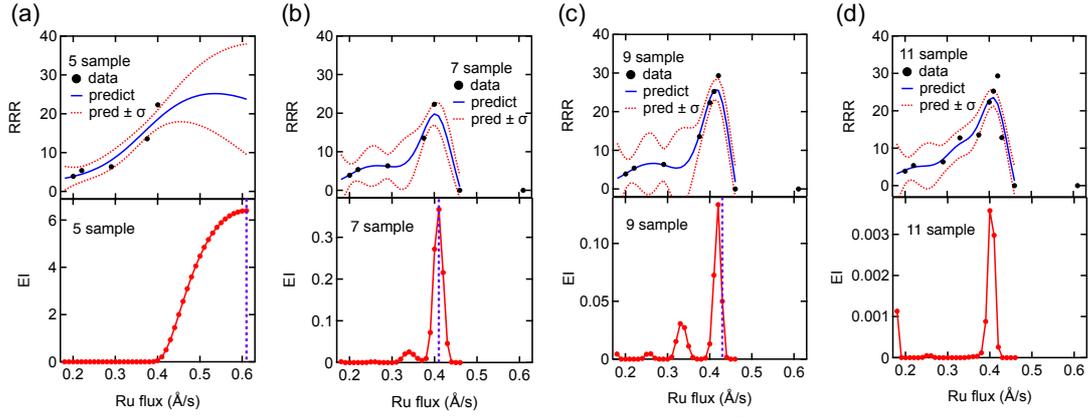

**FIG. 2.** Experimental and predicted RRR values with (a) five random initial Ru flux rates and with (b) seven, (c) nine, and (d) eleven Ru flux rates determined by the BO algorithm. Here, the EI values are also shown in the lower panels. The black filled circles, blue solid curves, and red dashed curves represent the experimental RRR, predicted RRR, and predicted RRR ±σ, respectively. In (a)-(c), the purple dashed lines represent the next sampling points.

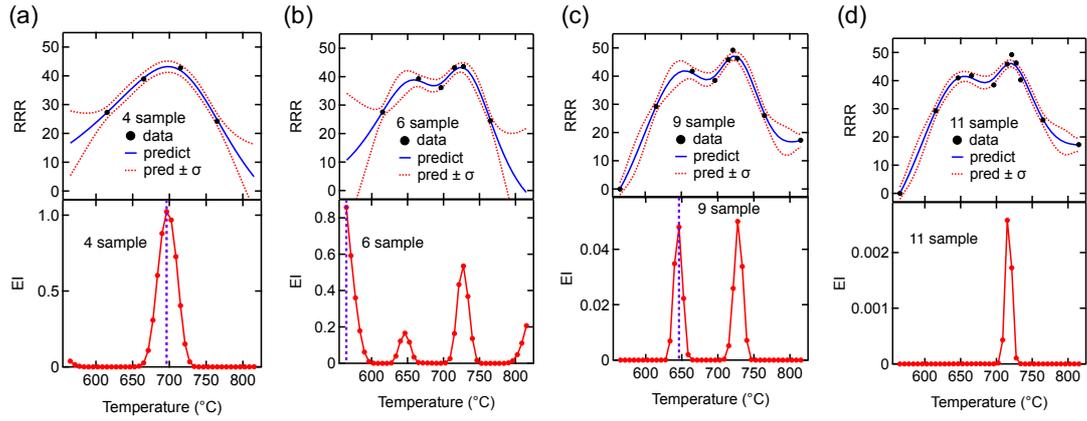

**FIG. 3.** Experimental and predicted RRR values with (a) four initial growth temperatures and with (b) six, (c) nine, and (d) eleven growth temperatures determined by the BO algorithm. Here, the EI values are also shown in the lower panels. The black filled circles, blue solid curves, and red dashed curves represent the experimental RRR, predicted RRR, and predicted RRR ±σ, respectively. In (a)-(c), the purple dashed lines represent the next sampling points.

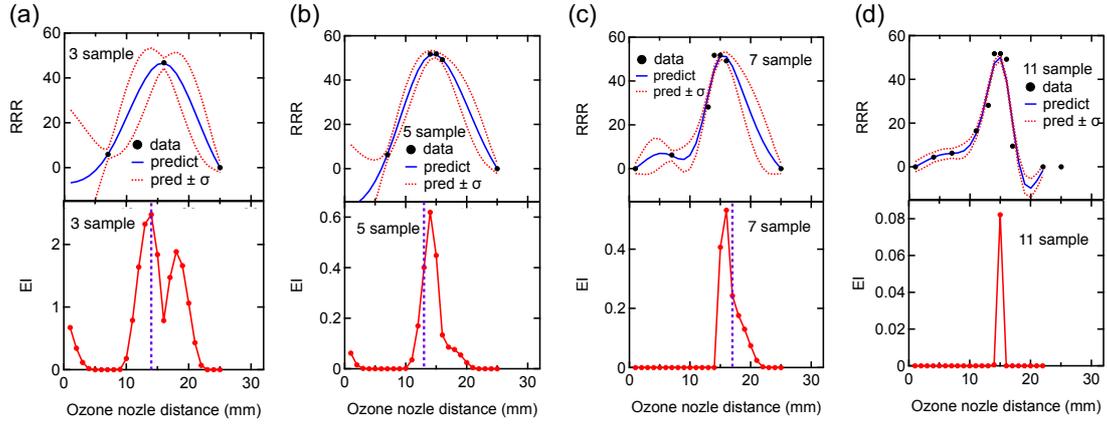

**FIG. 4.** Experimental RRR and predicted RRR values with (a) three initial $O_3$-nozzle-to-substrate distances and with (b) five, (c) seven, and (d) eleven $O_3$-nozzle-to-substrate distances determined by the BO algorithm. Here, the EI values are also shown in the lower panels. The black filled circles, blue solid curves, and red dashed curves represent the experimental RRR, predicted RRR, and predicted RRR $\pm\sigma$, respectively. In (a)-(c), the purple dashed lines represent the next sampling points.

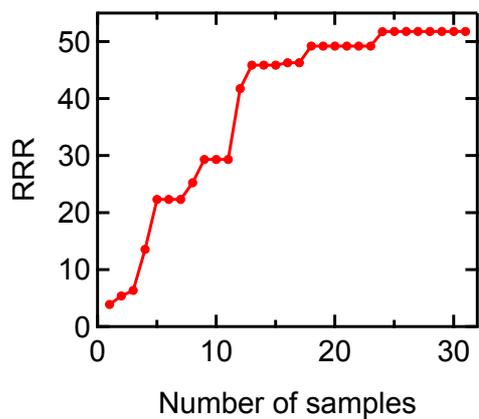

**FIG. 5.** Highest experimental RRR values plotted as a function of the total number of MBE growth runs.

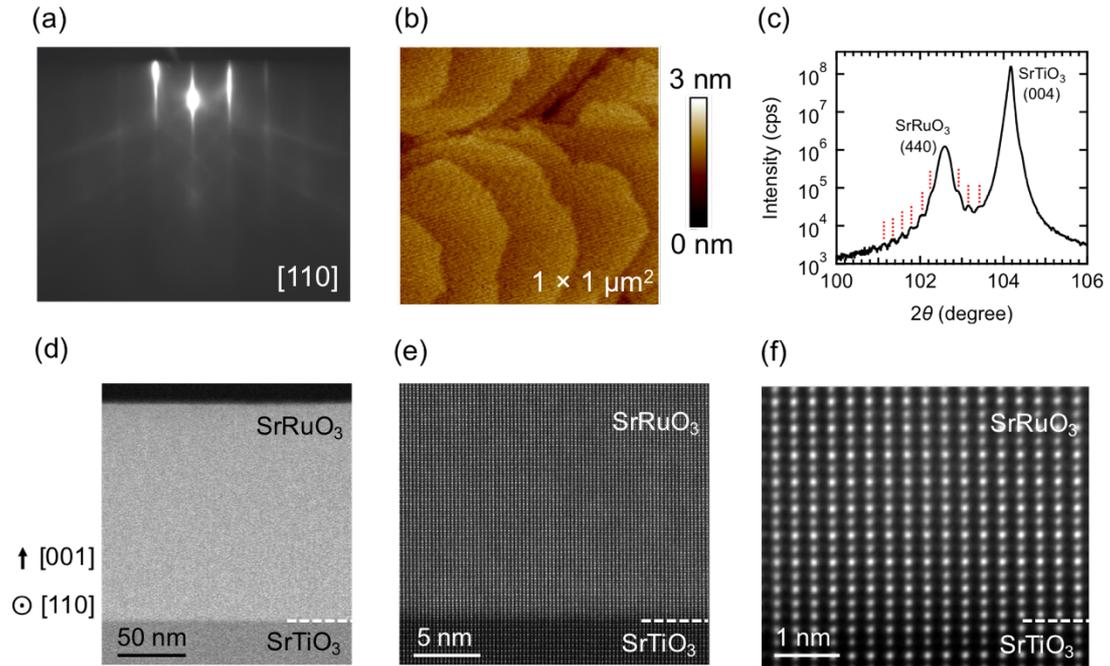

**FIG. 6.** (a) RHEED pattern (b) AFM image, (c) $\theta$-$2\theta$ scanned XRD pattern, and (d) cross-sectional HAADF-STEM image of a SrRuO$_3$ thin film with the RRR of over 50. (e) Magnified image near the interface in (d). (f) Magnified image near the interface in (e). In (a), the incident electron beam is parallel to the [110] axis of the SrTiO$_3$ substrate. In (b), the scan area is 1 × 1 μm$^2$. In (d)-(f), images are taken along the [110] direction of the SrTiO$_3$ substrate.

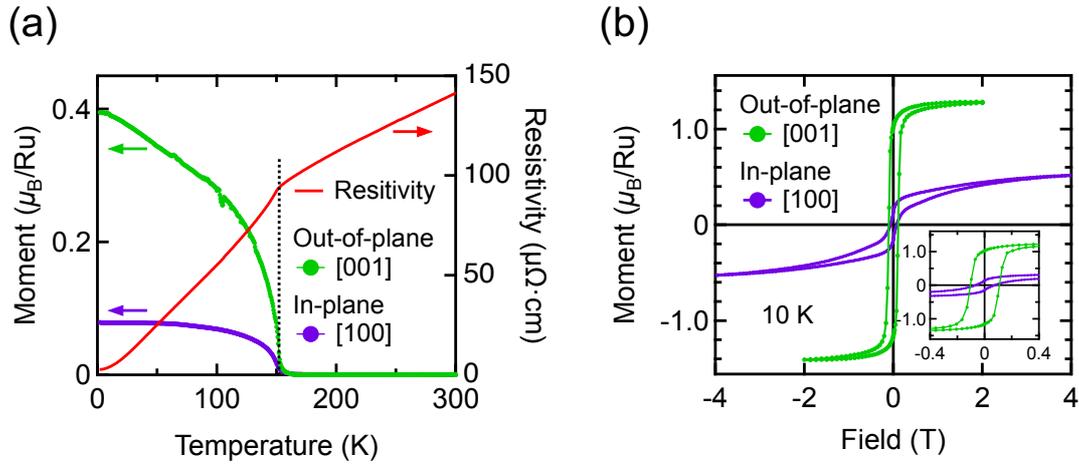

**FIG. 7.** (a) Resistivity versus temperature curve (red curve) and magnetization versus temperature curves of a SrRuO$_3$ film with the RRR of over 50. The magnetization versus temperature curves were measured in a field-cooled cooling process with a magnetic field of 100 Oe applied to the out-of-plane [001] (green filled circles) and in-plane [100] (purple filled circles) directions of the SrTiO$_3$ substrate. In (a), the black dashed line indicates $T_C$. (b) Magnetization versus magnetic field curves at 10 K with magnetic fields applied to the out-of-plane [001] (green circles) and in-plane [100] (purple circles) directions of the SrTiO$_3$ substrate.